\documentclass[conference]{IEEEtran}
\IEEEoverridecommandlockouts

\usepackage{standalone}
\usepackage[reqno]{amsmath}
\usepackage{amssymb}

\usepackage{algorithm}
\usepackage{algorithmicx}
\usepackage{algpseudocode}
\usepackage{setspace}
\usepackage{multicol}
\usepackage{enumerate}
\usepackage[font=small, belowskip=-15pt,aboveskip=0pt]{caption}
\usepackage{cite}
\usepackage{amsthm}
\usepackage[]{graphicx}
\usepackage{tikz}
\usepackage{tkz-tab}
\usepackage{pgfplots}
\usepackage{booktabs}
\usepackage{floatflt}

\usepackage{psfrag}	
\usepackage{array}
\usepackage{multirow,hhline}
\usepackage{exscale}
\usepackage{color}
\usepackage{colortbl}
\usepackage{epsfig}

\usepackage{array}
\usepackage[tight]{subfigure}

\usepackage[latin1]{inputenc} 
\usepackage[T1]{fontenc} 
\usepackage{rotating}
\usepackage{pbox}

\usepackage{pifont}

\usepackage{comment}
\usepackage{bookmark}
\usepackage{cleveref}

\renewcommand{\vec}[1]{\ensuremath{\boldsymbol{#1}}}

\newtheorem{remark}{Remark}

\usetikzlibrary{shapes,snakes}
\usetikzlibrary{calc}
\usetikzlibrary{patterns}
\usetikzlibrary{decorations.pathmorphing} 
\usetikzlibrary{spy}
\usetikzlibrary{positioning}
\usetikzlibrary{arrows, decorations.markings}
\usetikzlibrary{shapes.multipart, shapes.geometric, fit, backgrounds}
\usetikzlibrary{spy}

\definecolor{mycolor1}{rgb}{0.00000,0.44700,0.74100}%
\definecolor{mycolor2}{rgb}{0.85000,0.32500,0.09800}%
\definecolor{mycolor4}{rgb}{0.92900,0.69400,0.12500}%
\definecolor{mycolor3}{rgb}{0.49400,0.18400,0.55600}%
\definecolor{mycolor5}{rgb}{0.46600,0.67400,0.18800}%
\definecolor{mycolor6}{rgb}{0.30100,0.74500,0.93300}%
\definecolor{mycolor7}{rgb}{0.63500,0.07800,0.18400}%

\begin{document}

\title{Rate-Splitting Enabled Multi-Connectivity in Mixed-Criticality Systems}

\author{\IEEEauthorblockN{Yasemin Karacora, Aydin Sezgin}
    \IEEEauthorblockA{ Institute of Digital Communication Systems, Ruhr University Bochum, Germany \\ Emails: \{yasemin.karacora, aydin.sezgin\}@rub.de \vspace{-.5cm}}
    \thanks{This work has been submitted to the IEEE for possible publication.
Copyright may be transferred without notice, after which this version may
no longer be accessible.}
\thanks{This work was funded by the Ministry of Economic Affairs, Industry, Climate Action and Energy of the State of North Rhine-Westphalia, Germany under grant 005-2108-0028 (5G-Furios).}}

\maketitle

 \begin{abstract}
The enormous quality of service (QoS) demands posed by mission-critical use-cases of future 5G/6G wireless communication raise the need for resource-efficient highly reliable and low latency connectivity solutions. Multi-connectivity is considered a promising yet resource demanding approach to enhance reliability. In this work, we study the potential of the rate-splitting multiple access (RSMA) framework as an efficient way to enable uplink multi-connectivity for data transmissions with particularly high reliability requirements. 
Mapping high-criticality data onto the common stream allows it to be decoded at multiple access points (APs), which enhances reliability, while the private stream is utilized to serve applications with less stringent requirements. We propose a criticality-aware RSMA-based transmission scheme with short blocklength coding and derive an iterative power allocation algorithm by means of successive convex approximation (SCA). The proposed scheme is shown to achieve an expanded stability rate region compared to two baseline schemes. Moreover, it turns out to be less impacted by short blocklength while leading to substantial rate gains, particularly in the high SNR regime.
\end{abstract}

\section{Introduction}
One of the main use-cases of the 5G and 6G mobile communication standard is mission- and safety-critical applications, such as industrial automation, V2X-communication or tactile internet, which require ultra reliable and low latency communication (URLLC).
Industrial applications, for instance, need to guarantee a latency of 1-10 ms and an error rate of $10^{-5} - 10^{-9}$ \cite{siddiqi20195g}.
This is a challenging task as in a dynamic environment, links may become temporarily unavailable due to blockage and deep fading, which jeopardizes communication reliability. In addition, as short blocklength coding is essential for URLLC applications to comply with the low-latency demands, we are facing a rate-reliability tradeoff \cite{zhao2022queue}. Communication outages will induce retransmissions or even handovers and thereby cause time delays. A promising approach to enable ultra-reliable communication despite link blockage, channel fading and short blocklengths, yet avoid handover interruption time delays, is the implementation of multi-connectivity \cite{suer2019multi}. However, transmitting redundant packets to multiple APs comes with an increase in resource consumption \cite{mahmood2012resource}. This raises the need for efficient multi-connectivity schemes that are able to achieve the stringent reliability specifications with limited resources.
In fact, many use-cases of 5G URLLC involve simultaneous transmission of heterogeneous data with diverse QoS demands \cite{jin2021mixed}. For instance, in an industrial maintenance scenario, video data for mixed-reality remote assistance requires high data rates in addition to URLLC, yet control commands or monitoring of safety-critical sensor measurements have more stringent latency and reliability requirements and should be prioritized. 
Hence, taking the different levels of criticality into account can lead to enhanced efficiency of the multi-connectivity approach, e.g., such that multiple communication paths are provided only for specific safety-critical services. 

To this end, we propose a novel criticality-aware uplink transmission and resource allocation scheme based on the concept of rate-splitting \cite{mao2022rate, xu2022rateFBL}.
In RSMA, the transmit data is split into a common and a private message stream and decoded successively in order to reduce interference for the private messages. RSMA has been shown to be a promising technology in 6G, as it is capable of not only increasing data rates, but also enhancing reliability in URLLC \cite{dizdar2021rate}. In contrast to conventional RSMA, we exploit the fact that common messages are decoded at multiple receivers as a resource-efficient way to establish multi-connectivity and thereby enhance reliability of the common stream.

\subsection{Related Work}
The potential of multi-connectivity as an enabler for URLLC has been addressed in recent works, e.g., in \cite{suer2019multi, barbarossa2019resilient, liu2020resource}, and references therein. The authors in \cite{barbarossa2019resilient} investigate the benefits of using concurrent links in Mobile Edge Computing (MEC) to reduce power consumption and prevent outages caused by channel blockage. In \cite{liu2020resource}, resource management is studied in a heterogeneous network, where multi-connectivity is shown to play a key role for guaranteeing reliability of the URLLC users. Moreover, the authors propose a non-orthogonal coexistence scheme for URLLC and eMBB with efficient spectrum reuse.
Resource management for users with diverse QoS demands is also considered in \cite{kalor2019ultra, park2020centralized, liu2022network, reifert2022comeback, reifert2022MC_RSMA}. The work in \cite{kalor2019ultra} studies ultra-reliable communication in combination with heterogeneous latency requirements. It is shown that non-orthogonal resource allocation for mixed-criticality users enhances reliability and resource efficiency. The authors in \cite{park2020centralized} consider Age-of-Information as a metric for timeliness and define different non-linear aging functions depending on the application type, thereby capturing the heterogeneity of the system. In \cite{liu2022network, reifert2022comeback, reifert2022MC_RSMA}, RSMA is used to address the resource management in mixed-criticality networks. The work in \cite{liu2022network} proposed a RSMA-based network slicing scheme to improve performance with coexistence of URLLC, eMBB and mMTC users. 
In \cite{reifert2022comeback}, the authors introduce a resilience metric that is used to develop a RSMA-based criticality-aware resource management algorithm. They further study the tradeoff between maintaining energy-efficiency and complying with the QoS target rates of different users in a mixed-criticality network in \cite{reifert2022MC_RSMA}.
Although \cite{kalor2019ultra, park2020centralized, liu2022network, reifert2022comeback, reifert2022MC_RSMA} consider heterogeneous QoS requirements, each user only demands one type of service. In contrast, in this work we assume that each user is generating data of different criticality levels. 

\subsection{Contribution}
In this work, we propose a RSMA-based mixed-criticality uplink transmission scheme that makes use of the multi-connectivity naturally occuring in RSMA for reliability enhancement of high-criticality data transmissions. 
To the best of our knowledge, this is the first paper studying the multi-connectivity potential of uplink RSMA. We consider a system comprising two users and two APs, where data packets of two different criticality levels are generated at each user to be transmitted to any of the APs. We formulate a power allocation optimization problem based on RSMA with finite blocklength (FBL), which is a non-convex fractional program. We then apply a first-order Taylor approximation and adopt a quadratic transform to obtain a convex approximation and solve the problem iteratively. Moreover, the effective error rate of our scheme is analyzed and compared to two baseline schemes. Eventually, we evaluate the performance of our proposed scheme by means of simulations. We identify a rate region, where the system can be stabilized, and which we show is substantially larger in our scheme compared to both benchmark schemes. The impact of blocklength and signal-to-noise ratio (SNR) on the rate gain of our scheme is studied. The relative rate loss induced by short blocklength is shown to be less significant with our proposed method. 

\section{System Model}
We consider the uplink of a simple wireless network comprising two single-antenna user equipments (UEs) and two APs. Mixed-criticality data is generated at each UE to be transmitted to the APs. For instance, when considering an industrial maintenance use-case, video data and sensor measurements could be generated for remote assistance as well as for safety monitoring purposes, leading to different requirements regarding data rates, latency and reliability. The mixed-criticality aspect is modeled in the form of two buffers at each UE, one for high-criticality (HC) and low-criticality (LC) data packets.
Note that this model follows the 5G New Radio (NR) QoS-flow framework, where packets are classified depending on their QoS requirements before being mapped to data radio bearers \cite{jung2020intelligent}. 
\begin{figure}[t]
    \centering
    \includegraphics[]{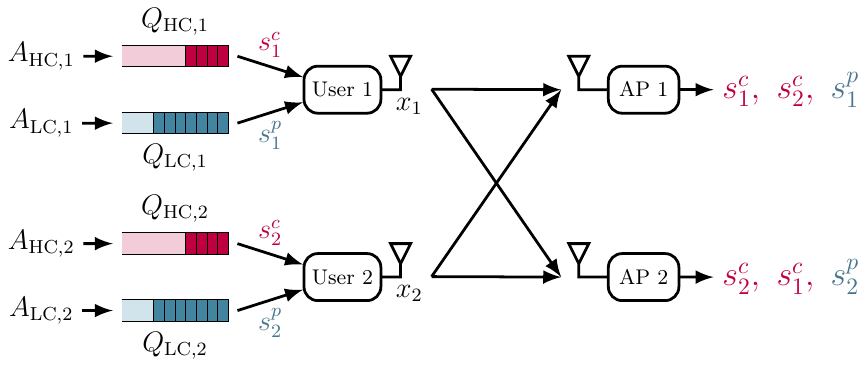}
    \caption{System model and RSMA-based transmission scheme}
    \label{fig:sys_model}
\end{figure}

\subsection{Rate-Splitting with Finite Blocklength Coding}
We introduce a RSMA-based transmission scheme to simultaneously transmit data of different criticality to multiple APs. 
The system model and transmission scheme is illustrated in Figure \ref{fig:sys_model}. 
User $i$ separately encodes the HC and LC messages to form a common and a private data stream, namely $s^c_i$ and $s^p_i$, respectively. Note that all HC packets are encoded as common messages, in order to allow them to be decoded at both APs and thereby increase the probability of successful delivery. Meanwhile, a private stream comprising LC data is decoded at one designated AP only. 
The mixed-criticality transmit signal of user $i$ is then given as: 
\begin{equation}
    \vec{x}_i =  \sqrt{p^c_{i}} s^c_{i} + \sqrt{p^p_{i}} s^p_i, \qquad i \in \{1,2\},
\end{equation}
where the transmit power of the common and private signal is denoted by $p^c_i$ and $p^p_i$, respectively.
The received signal at AP $i$ is hence obtained as
\begin{equation}
    y_i = h_{ii} x_i + h_{ij} x_j + n_i, \qquad i,j \in \{1,2\},~j \neq i,
\end{equation}
in which $h_{ij}$ represents the channel from user $j$ to AP $i$ and $n_i \sim \mathcal{CN}(0,{\sigma_i}^2)$ is additive white Gaussian noise (AWGN). We assume that $|h_{ii}| > |h_{ji}|$.
The APs apply a successive decoding (SD) strategy to decode both user's HC (common) messages and one UE's LC (private) message. More precisely, AP $i$ is assumed to decode three data streams in the following order: $s^c_i \rightarrow s^c_j \rightarrow s^p_i$. 
As a result, the signal-to-interference-plus-noise ratio (SINR) for these signals can be expressed as
\vspace{-2mm}
\begin{align}
    \Gamma^c_{ii} &= \frac{|h_{ii}|^2 p^c_i}{|h_{ii}|^2 p^p_i + |h_{ij}|^2 (p^c_j+p^p_j) + \sigma_i^2}, \label{gamma_c_ii}\\
    \Gamma^c_{ij} &= \frac{|h_{ij}|^2 p^c_j}{|h_{ii}|^2 p^p_i + |h_{ij}|^2 p^p_j + \sigma_i^2}, \label{gamma_c_ij}\\
    \Gamma^p_{ii} &= \frac{|h_{ii}|^2 p^p_i}{|h_{ij}|^2 p^p_j + \sigma_i^2}. \label{gamma_p_ii}
\end{align}
As we are considering URLLC applications, short blocklength coding is needed to meet the latency requirements. Hence, we utilize the FBL coding framework proposed in \cite{polyanskiy2010channel} for our achievable rate analysis. Let $l$ be the blocklength and $\epsilon$ represent the block error rate (BLER). Thus, the achievable rate is approximated as
\begin{equation} \label{rate_FBL}
    R(\Gamma,\epsilon) \approx B \left(\log_2(1 + \Gamma) - \log_2(e)\sqrt{\frac{V(\Gamma)}{l}} Q^{-1}(\epsilon) \right).
\end{equation}
Here, $B$ denotes the bandwidth, $V(\Gamma) = \left(1-(1+\Gamma)^{-2}\right)$ represents the channel dispersion, and $Q(x) = \frac{1}{\sqrt{2\pi}} \int_x^\infty \exp({-\frac{u^2}{2}}) \mathrm{d}u$ is the Q-function.
In order to guarantee that the common messages can be fully decoded and removed from the receive signal when applying the decoding order given above, the coding rate of the $i$-th UE's common message needs to meet the following constraint:
\begin{equation}
    R^c_i \leq \min\left\{R(\Gamma^c_{ii}, \epsilon_c), R(\Gamma^c_{ji},\epsilon_c)\right\}, \quad i,j \in \{1,2\}, i \neq j.
\end{equation}
Note that this condition enables multi-connectivity for the common message stream and thereby increases reliability of the HC data transmission. 
\vspace{-2mm}
\subsection{Queuing Model}
 The information packets of size $M$ bits generated at each UE are assumed to be classified according to their criticality level and buffered in two separate queues.  
 Assume that HC and LC data packets of size $M$ bits arrive at user $i$ with arrival rate $\bar{A}_{\mathrm{HC},i}$ and $\bar{A}_{\mathrm{LC},i}$, respectively. 
 Time is devided into slots of length $T$.
In each time slot, multiple packets of the same criticality are concatenated to a frame and are jointly encoded using short blocklength coding. 
Let $Q_{\mathrm{HC},i}(t)$ be the backlog of the $i$-th user's HC buffer at time slot $t$, which evolves as
\begin{equation}
    Q_{\mathrm{HC},i}(t+1) = \left[Q_{\mathrm{HC},i}(t) - \frac{T}{M} R^c_i \right]^+ + A_{\mathrm{HC},i}(t).
\end{equation}
Similarly, the LC queue length's evolution is obtained by
\begin{equation}
    Q_{\mathrm{LC},i}(t+1) = \left[Q_{\mathrm{LC},i}(t) - \frac{T}{M} R^p_i \right]^+ + A_{\mathrm{LC},i}(t).
\end{equation}
In order to prevent buffer overflow, the transmission scheme should be designed in a way that all queues are stabilized. A queue is denoted as stable, if 
\begin{equation}
    \lim_{t\rightarrow \infty} \sup \frac{1}{t} \sum_{\tau=0}^{t-1} \mathbb{E} [Q(\tau)] < \infty.
\end{equation}
Note that since all HC packets function as common messages to ensure their reliability, the ratio of common versus private data rate must be dependent on the HC and LC arrival rates for queue stability. This is a major difference between our proposed scheme and conventional RSMA.

\section{Problem Formulation and Proposed Scheme}
Our goal is the stabilization of the queues while guaranteeing the reliability requirements of the mixed-criticality applications. Hence, we formulate an optimization problem to determine the power allocated to the different data streams. 
In order to stabilize the queues at both UEs and prevent buffer overflow, the average rate of successfully transmitted data has to be larger than the arrival rate for each queue.
The optimization problem is considered for a fixed blocklength $l$ and BLERs $\epsilon_c$ and $\epsilon_p$ for the common and private messages, respectively. For readability, we introduce the vectors $\vec{p} = [p^c_1, p^c_2, p^p_1, p^p_2]$ and $\vec{R} = [R^c_1, R^c_2, R^p_1, R^p_2]$.
Furthermore, let $f(\vec{R})$ be a concave function, which serves as maximization objective. Thus, we consider the following optimization problem:
\begin{subequations} \label{opt_original}
\begin{align}
\max_{\vec{p}, \vec{R}} ~&f(\vec{R}) \tag{\theparentequation}\vspace{-2mm}\\
\text{s.t.}~~ &\bar{A}_{\mathrm{HC},i} - \frac{T}{M} R^c_i \leq 0, \quad i \in \{1,2\}, \label{HC_stability_constr}\\
 &\bar{A}_{\mathrm{LC},i} - \frac{T}{M} R^p_i \leq 0, \quad i \in \{1,2\}, \label{LC_stability_constr} \\
&r^c_{ij} \leq R\left(\Gamma^c_{ij}, \epsilon_c\right), \quad i,j \in\{1,2\}, \label{c_rate_constr}\\
&R^c_i \leq \min\{r^c_{ii}, r^c_{ij}\}, \quad i,j \in \{1,2\},~~j \neq i, \label{RSMA_constr}\\
&R^p_{i} \leq R\left(\Gamma^p_{ii},\epsilon_p\right), \quad i \in \{1,2\}, \label{p_rate_constr}\\
 & p^c_i + p^o_i \leq P^\mathrm{max}_i, \quad i \in \{1,2\}. \label{power_constr}
\end{align}
\end{subequations}
In \eqref{opt_original}, the constraints \eqref{HC_stability_constr} and \eqref{LC_stability_constr} ensure stability of the queues while \eqref{RSMA_constr} allows the common messages to be decoded at both APs. Note that \eqref{opt_original} is a non-convex problem due to the non-convex constraints \eqref{c_rate_constr} and \eqref{p_rate_constr}.
We utilize a first-order Taylor expansion of the FBL rate terms as well as a quadratic transform to tackle the fractional SINR-expressions to derive a convex approximation of problem \eqref{opt_original}. In this manner, the power allocation problem can be efficiently solved using SCA. The detailed problem reformulation as well as the iterative optimization algorithm are provided in Appendix \ref{appendixA}.
\begin{remark}
We do not specify the objective function $f(\vec{R})$ in this section. Suitable objectives could be a Lyapunov drift function to improve queue stabilization, or the weighted sum-rate. Since this paper focuses on fundamentally assessing the potential of the proposed RSMA-based criticality-aware transmission scheme, we are evaluating the maximum achievable LC rates for fixed HC rates in the simulation section. A study of the queue backlog over time is reserved for future work.
\end{remark}

\section{Reliability Analysis and Baseline Schemes}
As a consequence of the successive decoding, the effective error probability for each message differs from $\epsilon$ in \eqref{rate_FBL} as the decoding errors of the messages decoded first propagate to the lastly decoded message. 
Assume that the common and private messages of UE $i$ are encoded with blocklength $l$ and BLER $\epsilon^c_i$ and $\epsilon^p_i$, respectively. While the common message $s^c_i$ is the first message being decoded at AP $i$ with BLER $\epsilon^c_i$, it can be successfully decoded at AP $j$ only if the decoding of $s^c_j$ has been successful as well. The common message is lost if both APs fail at decoding it. Thus, the effective error rate of the common message $q^c_i$ can be obtained as:
\begin{equation} \label{q_c}
    q^c_i = \epsilon^c_i \left(1 - (1-\epsilon^c_i)(1-\epsilon^c_j)\right).
\end{equation}
Since the private message is decoded at only one of the APs after both common messages were removed from the receive signal, it's decoding requires the successful decoding of both common messages. Hence, the effective error probability of $s^p_i$ is given by:
\begin{equation} \label{q_p}
    q^p_i = 1 - (1-\epsilon^c_i)(1-\epsilon^c_j)(1-\epsilon^p_i).
\end{equation}
For $\epsilon^c_1 = \epsilon^c_2=\epsilon_c$ and $\epsilon^p_1 = \epsilon^p_2=\epsilon_p$ and given that $\epsilon_c, \epsilon_p \ll 1$, the effective error probabilities for HC and LC data streams can be approximated from \eqref{q_c} and \eqref{q_p} as $q_c \leq 2 {\epsilon_c}^2$ and $q_p \leq 2 \epsilon_c + \epsilon_p$, respectively. Indeed, due to multi-connectivity the effective error rate of the common message is significantly lower than the BLER of a single link. However, the effective error rate of the private message increases due to error propagation and the SD strategy.
As a consequence, assuming that $q_\mathrm{HC}$ and $q_\mathrm{LC}$ are given QoS requirements for the reliability of HC and LC data packets, the BLER that should be taken into account when selecting the transmission rate can be determined as\footnote{In this work, we assume that reliability is only affected by the BLER induced by FBL coding. This restriction allows for a fair comparison of the rate-reliability tradeoff in multi- and single-connectivity scenarios. However, deep fading and channel blockage substantially impact the reliability and are prevalent in dynamic environments. Establishing multi-connectivity is therefore essential to guarantee high reliability, yet avoid large delays due to handover interruption time.}:
\begin{align}
    \epsilon_c &= \sqrt{\frac{q_\mathrm{HC}}{2}}, \label{BLER_HC} \\
    \epsilon_p &= q_\mathrm{LC} - \sqrt{2 q_\mathrm{HC}}.
\end{align}

\subsection{TDM and SC baseline schemes}
We compare our proposed RSMA-based scheme with two baselines, which we refer to as multi-connectivity (MC) and single-connectivity (SC) Time-Division Multiplexing (TDM). In this way, we investigate the benefits of our scheme gained from multi-connectivity on the one hand, and the non-orthogonal superposition coding approach on the other hand. 
In both reference schemes, the HC and LC data is transmitted successively by splitting each time slot into a HC phase and a LC phase. Hence, for queue stability the rates need to satisfy
\begin{align}
    \bar{A}_{\mathrm{HC},i} &\leq \alpha \frac{T}{M} R^c_i, \label{HC_stab_TDM}\\
    \bar{A}_{\mathrm{LC},i} &\leq (1-\alpha) \frac{T}{M} R^p_i,  \label{LC_stab_TDM}  
\end{align}
where $\alpha \in [0,1]$ determines the fraction of each time slot reserved for the HC data transmission. 
\paragraph{SC-TDM} 
In SC-TDM, only the associated UE's message is decoded at each AP in both phases while interference from the other UE is simply treated as noise. 
Thus, the optimal power allocation is determined by solving the following optimization problem:
\begin{subequations}
\begin{align}\label{opt_SC_TDM}
    \max_{\vec{p},\vec{R}, \alpha}~ &f(\vec{R}) \tag{\theparentequation}\\
    \text{s.t.}~~ & \eqref{HC_stab_TDM}, \eqref{LC_stab_TDM}, \nonumber \\
    & R^c_i \leq R\left(\frac{p^c_i|h_{ii}|^2}{p^c_j |h_{ij}|^2 + \sigma_i^2}, ~q_\mathrm{HC}\right), \label{HC_rate_SC-TDM}\\
    & R^p_i \leq R\left(\frac{p^p_i|h_{ii}|^2}{p^p_j |h_{ij}|^2 + \sigma_i^2},~ q_\mathrm{LC}\right), \label{LC_rate_SC-TDM}\\
    & p^c_i, p^p_i \leq P^\mathrm{max}_i, \quad i \in \{1,2\}. \label{power_constr_TDM}
\end{align}
\end{subequations}
Due to the single-connectivity, in order to provide the same reliability as our proposed scheme, i.e., effective error probabilities $q_c$ and $q_p$, the achievable rates during the HC and the LC phase are limited by \eqref{HC_rate_SC-TDM} and \eqref{LC_rate_SC-TDM}, respectively. Since the only difference between transmission in the HC and LC phase is the BLER requirement, the optimal transmit powers are the same in both phases.
\paragraph{MC-TDM}
In contrast to SC-TDM, the APs successively decode the critical messages from both UEs during the HC phase in the MC-TDM scheme, while in the LC phase only the designated UE's message is decoded. Thereby, multi-connectivity is maintained for the HC messages, though interference of the private messages in the LC phase is treated as noise. Thus, the corresponding optimization problem can be written as
\begin{subequations}
\begin{align} \label{opt_MC_TDM}
    \max_{\vec{p},\vec{R}, \alpha}~ &f(\vec{R}) \tag{\theparentequation}\\
    \text{s.t.}~~ & \eqref{HC_stab_TDM}, \eqref{LC_stab_TDM}, \eqref{LC_rate_SC-TDM}, \eqref{power_constr_TDM}, \nonumber \\
    & R^c_i \leq R\left(\min\left\{\frac{p^c_i|h_{ii}|^2}{p^c_j |h_{ij}|^2 + \sigma_i^2}, \frac{p^c_i |h_{ji}|^2}{\sigma_i^2}\right\},~ \sqrt{q_\mathrm{HC}/2}\right), \label{HC_rate_MC-TDM}
\end{align}
\end{subequations}
in which \eqref{HC_rate_MC-TDM} ensures that the HC messages can be decoded at both APs and the BLER for transmission is chosen as in \eqref{BLER_HC} due to multi-connectivity. 
Note that both problems \eqref{opt_SC_TDM} and \eqref{opt_MC_TDM} can be solved iteratively by applying the SCA-method described in Appendix \ref{appendixA}.

\section{Numerical Results}
We evaluate the performance of our proposed criticality-aware RSMA-scheme compared to the SC- and MC-TDM reference schemes. For simplicity, we consider a symmetric AWGN-channel consisting of two UEs and two APs. We define $h_{11}=h_{22} = 1$ and $h_{12} = h_{21} = 0.6$. The SNR is $P^\mathrm{max}_i/\sigma_i^2=15$ dB, $i\in\{1,2\}$. The reliability requirements for the HC and LC data is set to $q_\mathrm{HC} = 10^{-7}$ and $q_\mathrm{LC}=10^{-3}$, respectively, while the blocklength is assumed to be $l=1000$ if not stated otherwise. The bandwidth is $B = 300$ kHz, the duration of a time slot $T = 5$ ms, and a packet contains $M=128$ bits.
For simplicity of the performance evaluation, we consider a symmetric system with equal parameters for the two users. Hence, we evaluate the achievable rates of one user only, as there is no difference between UE 1 and UE 2.
\begin{figure*}[]
    \centering
    \subfigure[Stability region boundary]{\label{fig:stability_region}
   \includegraphics[]{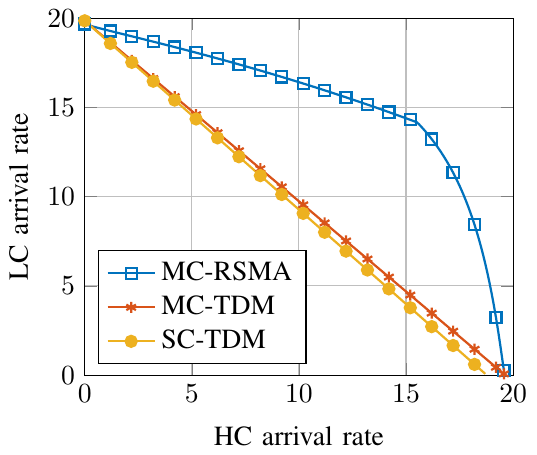}}
    \subfigure[Relative LC rate loss induced by FBL]{\label{fig:rate_BL}
   \includegraphics[]{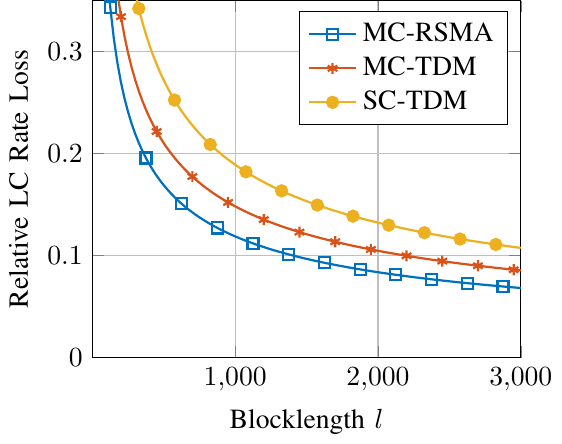}}
    \subfigure[Achievable LC rate vs. SNR]{\label{fig:rate_vs_snr}
    \includegraphics[]{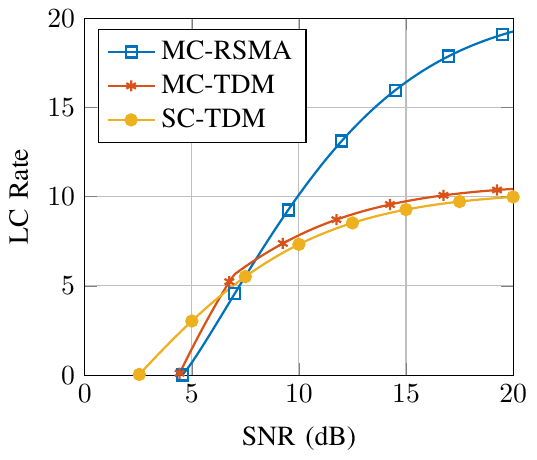}}
    \caption{Achievable stability region and impact of blocklength and SNR on the achievable LC rate for $\Bar{A}_{\mathrm{HC}}=10$ packets/slot}
    \label{fig:LC_rate}
\end{figure*}
\subsection{Stability Region}
%
%
First, we identify the stability region of our scheme, i.e., the region of HC and LC arrival rates for which the queues can be stabilized \cite{boche2004stability}. The boundary of this region is obtained by maximizing $\min\{\Bar{A}_{\mathrm{LC},1}, \Bar{A}_{\mathrm{LC},2}\}$ for a fixed HC arrival rate $\Bar{A}_{\mathrm{HC},i}$ to maintain user fairness. Under the assumptions made above, we can equivalently set $f(\vec{R}) = \min\{R^p_1, R^p_2\}$ in Problem \eqref{opt_original}, \eqref{opt_SC_TDM}, and \eqref{opt_MC_TDM} and solve it as described in Appendix \ref{appendixA}. Note that in the considered symmetric system, it is optimal in the two TDM-schemes that both UEs transmit with full power the whole time. 

The achievable stability region boundary of our proposed scheme and the two baselines is shown in Figure \ref{fig:stability_region}.
Obviously, the proposed MC-RSMA scheme enables a significantly larger stability region than the two baseline schemes. This is for two reasons, namely the superposition coding and the multi-connectivity, which ensures the reliability of HC data despite transmitting at a higher rate. 
Both TDM-schemes do not apply superposition coding but only transmit one message at a time, splitting each time slot into a HC and a LC phase. Therefore, the stability region boundary is a linear function of the parameter $\alpha$. However, there is a gap between MC- and SC-TDM, which results from the multi-connectivity that MC-TDM enables during the HC phase. While in SC-TDM, each message is decoded at only one AP, both APs decode the HC messages in MC-TDM. Therefore, each transmission can have a BLER of $\sqrt{q_\mathrm{HC}/2}$ instead of $q_\mathrm{HC}$, which allows for a higher transmission rate. Notice that while the two schemes featuring multi-connectivity can stabilize the HC queue with an arrival rate of up to 19.6 packets/slot, only up to 18.7 packets/slot can be achieved with single-connectivity.

%
In Figure \ref{fig:rate_BL}, we study the relative rate loss that is caused by short blocklength in our proposed scheme as well as the baseline schemes. More precisely, for a fixed HC rate, we consider the percentage of the achievable LC rate, that is lost compared to the achievable rate with infinite blocklength. Note that from Fig. \ref{fig:stability_region}, it is evident, that MC-RSMA achieves higher rates compared to the baseline schemes. Fig. \ref{fig:rate_BL} additionally demonstrates, that our proposed scheme is also less affected by short blocklength than the TDM-schemes. Thus, a shorter blocklength could be selected in our scheme, which correlates with lower communication latency. 

Figure \ref{fig:rate_vs_snr} shows the maximum achievable LC rate with fixed HC rate for the SNR range from 0 to 20 dB. The proposed MC-RSMA scheme significantly outperforms the two baseline schemes for SNR greater than 8 dB. This is due to the superposition coding of HC and LC data streams and the successive decoding strategy at the APs, which allows the LC message to be transmitted at a low power. Below of 8 dB, however, the RSMA-based scheme is outperformed by the TDM-schemes. Both MC-RSMA and MC-TDM cannot achieve the HC rate of 10 packets/slot for SNR $< 4.4$ dB, because those schemes require the common messages to be decoded at both APs. In contrast, the SC-TDM fulfills the HC rate requirement with an SNR as low as 2.6 dB. 
Note that in conventional RSMA, where the ratio of common and private rate is flexible, the benefits of a common stream for rate maximization only take effect at higher SNR. In our proposed scheme, however, the common rate must at least stabilize the HC queue. Adding more flexibility to the criticality-aware assignment of messages to common and private streams could improve our scheme for a wide SNR range in future work.

\section{Conclusion}
This work studies the potential of RSMA as an efficient way of establishing multi-connectivity and thereby complying with strict, but heterogeneous reliability requirements of mission-critical 5G/6G applications. The concurrent decoding of common messages at multiple receivers, which is the core concept of the interference mitigation strategy in RSMA, is exploited as a way to enhance reliability of HC data transmission. In turn, data packets with less stringent reliability demands are transmitted within the private data stream. 
This paper considers a mixed criticality system comprising two UEs and two APs with HC and LC buffers. A queue stabilization problem is formulated, based on the RSMA framework and FBL coding. By means of a Taylor approximation and quadratic transform, a SCA-based power allocation algorithm is proposed. Simulation results show that the proposed MC-RSMA achieves a larger stability region than two TDM-based baseline schemes. Furthermore, the impact of short blocklength on the achievable rate is less detrimental with our scheme. MC-RSMA is shown to be especially beneficial in the higher SNR regime.

\appendix
 \subsection{SCA-based Solution}
\label{appendixA}
We adopt a first-order Taylor approximation to the square root of the channel dispersion as proposed in \cite{xu2022rateFBL, xu2022uplink}. With $D(\epsilon) = \log_2(e)\frac{Q^{-1}(\epsilon)}{\sqrt{l}}$, we obtain the following approximate rate expression around the point $\Tilde{\Gamma}$:
\begin{equation}\begin{split}
    \rho_{\Tilde{\Gamma}}(\Gamma) = & B\left(\log_2(1+\Gamma) \vphantom{\left[\frac{(1+\Tilde{\Gamma})^{-3}}{\sqrt{V(\Tilde{\Gamma})}}\right]} \right. \\ 
    &\left. -D(\epsilon) \left[\frac{(1+\Tilde{\Gamma})^{-3}}{\sqrt{V(\Tilde{\Gamma})}} (\Gamma - \Tilde{\Gamma}) + \sqrt{V(\Tilde{\Gamma})}\right] \right).
    \end{split}
\end{equation}
By introducing auxiliary variables $\vec{\gamma} = [\gamma^c_1, \gamma^c_2, \gamma^p_1, \gamma^p_2]$ we can approximate problem \eqref{opt_original} as:
\begin{subequations}
\begin{align}
    \max_{\vec{p},\vec{R},\vec{\gamma}} ~&f(\vec{R}) \tag{\theparentequation}\\
\text{s.t.}~~ &\eqref{HC_stability_constr}, \eqref{LC_stability_constr}, \eqref{RSMA_constr}, \eqref{power_constr}, \nonumber\\
&r^c_{ij} \leq \rho_{\Tilde{\Gamma}^c_{ij}}(\gamma^c_{ij}), \quad i,j,\in \{1,2\}, \label{c_rate_constr1}\\
&R^p_{i} \leq \rho_{\Tilde{\Gamma}^p_{ii}}(\gamma^p_{ii}), \quad i \in \{1,2\}, \label{p_rate_constr1}\\
 & \gamma^c_{ij} \leq \Gamma^c_{ij} , \quad i,j \in \{1,2\}, \label{gamma_c_constr}\\
 & \gamma^p_{ii} \leq \Gamma^p_{ii}, \quad i \in \{1,2\}. \label{gamma_p_constr}
\end{align}
\end{subequations}
This problem is still non-convex due to the fractional SINR-constraints. Hence, we utilize the quadratic transform proposed in \cite{shen2018fractional} and by introducing the auxiliary variables $\vec{\mu} = [\mu^c_{11},\mu^c_{12}, \mu^c_{21}, \mu^c_{22},  \mu^p_1, \mu^p_2]$, we obtain the functions
\begin{align}
\begin{split}
g^{c}_{ii}(\vec{p}, \vec{\gamma})& = \gamma^c_{ii} - 2 {\mu^c_{ii}} |h_{ii}|\sqrt{p^c_i}\\  &+ ({\mu^c_{ii}})^2 \left(|h_{ii}|^2 p^p_i + |h_{ij}|^2 (p^c_j+p^p_j) + \sigma_i^2\right),
\end{split} \label{g_c1}\\
\begin{split}
g^{c}_{ij}(\vec{p}, \vec{\gamma})& = \gamma^c_{ij} - 2 {\mu^c_{ij}} |h_{ij}|\sqrt{p^c_j}\\  &+ ({\mu^c_{ij}})^2 \left(|h_{ii}|^2 p^p_i + |h_{ij}|^2 p^p_j + \sigma_i^2\right),
\end{split} \label{g_c2}\\
\begin{split}
g^{p}_{i}(\vec{p}, \vec{\gamma})& = \gamma^p_{ii} - 2 {\mu^p_{i}} |h_{ii}|\sqrt{p^p_i}+ ({\mu^p_{i}})^2 \left(|h_{ij}|^2 p^p_j + \sigma_i^2\right). \label{g_p}
\end{split}
\end{align}
As derived in \cite{shen2018fractional}, when $\vec{p}$ and $\vec{\gamma}$ are kept fixed, the optimal auxiliary variables are obtained by setting the derivatives of \eqref{g_c1} -- \eqref{g_p} to zero, leading to:
\begin{align}
{\mu^c_{ii}}^* &= \frac{|h_{ii}| \sqrt{{p^c_i}^*}}{|h_{ii}|^2 {p^p_i}^* + |h_{ij}|^2 ({p^c_j}^*+{p^p_j}^*) + \sigma_i^2}, \label{mu_c1}\\
{\mu^c_{ij}}^* &= \frac{|h_{ij}| \sqrt{{p^c_j}^*}}{|h_{ii}|^2 {p^p_i}^* + |h_{ij}|^2 {p^p_j}^* + \sigma_i^2}, \label{mu_c2}\\
{\mu^p_{i}}^* &= \frac{|h_{ii}| \sqrt{{p^p_i}^*}}{|h_{ij}|^2 {p^p_j}^* + \sigma_i^2}. \label{mu_p}
\end{align}
Note that for fixed auxiliary variables $\vec{\mu}$, the quadratic transform functions \eqref{g_c1} -- \eqref{g_p} are convex functions in $\vec{p}$ and $\vec{\gamma}$. Hence, with the approximations described above, we obtain the following convex optimization problem for fixed $\vec{\mu}$ and $\Tilde{\Gamma}$:
\vspace{-2mm}
\begin{subequations}
\begin{align} \label{opt_convex}
\max_{\vec{p}, \vec{R},\vec{\gamma}}~ & f(\vec{R}) \tag{\theparentequation}\\
\text{s.t.}~~ &\eqref{HC_stability_constr}, \eqref{LC_stability_constr}, \eqref{RSMA_constr}, \eqref{power_constr}, \eqref{c_rate_constr1}, \eqref{p_rate_constr1}, \nonumber\\
& g^c_{ii}(\vec{p}, \vec{\gamma}) \leq 0,\quad i \in \{1,2\}, \\
 &g^c_{ij}(\vec{p},\vec{\gamma}) \leq 0, \quad i,j \in \{1,2\},~i\neq j,\\
 & g^p_{i}(\vec{p},\vec{\gamma}) \leq 0, \quad i \in \{1,2\}.
\end{align}
\end{subequations}
Finally, the power allocation problem is solved by SCA, i.e., solving a sequence of convex subproblems instead. In each iteration, problem \eqref{opt_convex} is solved by means of convex optimization solvers, such as CVX \cite{cvx}.
Then, the auxiliary variables $\vec{\mu}$ and the Taylor approximation point $\Tilde{\Gamma}$ are updated based on the optimized power allocation $\vec{p}$. The proposed iterative optimization algorithm is summarized in Alg. \ref{alg:rsma_alg}.
\begin{algorithm}[]
\caption{Power Allocation for MC-RSMA}
  \begin{algorithmic}[1]
    \State Initialize $\vec{p}$
    \Repeat 
        \State Compute ${\Tilde{\Gamma}^c_{ii}},~\Tilde{\Gamma}^c_{ij},~ \Tilde{\Gamma}^p_{ii}$ based on \eqref{gamma_c_ii} --  \eqref{gamma_p_ii}
        \State Compute $\mu^c_{ii},~\mu^c_{ij},~\mu^p_{i}$ based on \eqref{mu_c1} -- \eqref{mu_p}
        \State Solve \eqref{opt_convex}
    \Until{Convergence}
    \end{algorithmic}
    \label{alg:rsma_alg}
\end{algorithm}
\vspace{-3mm}

\bibliographystyle{./bibliography/IEEEtran}
\bibliography{./bibliography/IEEEabrv,./bibliography/IEEEexample,./bibliography/references}

\end{document}